\documentclass[runningheads,a4paper]{llncs}

\usepackage{cmap}
\usepackage[T1]{fontenc}

\usepackage{graphicx}
\usepackage{booktabs}
\usepackage{listings}
\usepackage{algorithmicx}
\usepackage{amsmath}
\usepackage{amsfonts}
\usepackage{amssymb}
\usepackage[breakall]{truncate}
\usepackage{caption} 
\usepackage{multirow}

\usepackage{url}
\def\UrlBreaks{\do\/\do-}
\usepackage{breakurl}

\usepackage[ngerman,english]{babel}
\addto\extrasenglish{\languageshorthands{ngerman}\useshorthands{"}}

\usepackage[%
rm={oldstyle=false,proportional=true},%
sf={oldstyle=false,proportional=true},%
tt={oldstyle=false,proportional=true,variable=true},%
qt=false%
]{cfr-lm}
\usepackage[math]{blindtext}

\usepackage{cite}

\usepackage{paralist}

\usepackage{csquotes}

\usepackage{microtype}

\usepackage{url}
\makeatletter
\g@addto@macro{\UrlBreaks}{\UrlOrds}
\makeatother

\usepackage[table]{xcolor}
\usepackage{tikz}

\usepackage[capitalise,nameinlink]{cleveref}
\crefname{section}{Sect.}{Sect.}
\Crefname{section}{Section}{Sections}

\usepackage{siunitx}
\usepackage{amsmath}
\usepackage{xspace} 
\usepackage{footnote}
\usepackage[]{algorithm}
\usepackage[noend]{algpseudocode}
\makesavenoteenv{tabular}
\makesavenoteenv{table}

\algrenewcommand{\algorithmiccomment}[1]{\hfill$\diamond$ #1}

\newcommand{\eg}{e.\,g.,\ }

\DeclareFontFamily{U}{MnSymbolC}{}
\DeclareSymbolFont{MnSyC}{U}{MnSymbolC}{m}{n}
\DeclareFontShape{U}{MnSymbolC}{m}{n}{
    <-6>  MnSymbolC5
   <6-7>  MnSymbolC6
   <7-8>  MnSymbolC7
   <8-9>  MnSymbolC8
   <9-10> MnSymbolC9
  <10-12> MnSymbolC10
  <12->   MnSymbolC12%
}{}
\DeclareMathSymbol{\powerset}{\mathord}{MnSyC}{180}

\newcommand\hmm[1]{\ifnum\ifhmode\spacefactor\else2000\fi>1000 \uppercase{#1}\else#1\fi}

\newcommand{\SC}[0]{\hmm{s}mart contract\xspace}
\newcommand{\SCS}[0]{\hmm{s}mart contracts\xspace}

\newcommand{\EP}[0]{Ethereum platform\xspace}
\newcommand{\ET}[0]{Ethereum\xspace}

\newcommand{\EVM}[0]{EVM\xspace}
\newcommand{\SOL}[0]{Solidity\xspace}
\newcommand{\EA}[0]{et al.\xspace}
\newcommand{\ETC}[0]{etc.\@\xspace}

\newcommand{\ERC}[0]{ERC-20\@\xspace}

\newcommand{\TOKS}[0]{token systems\@\xspace}

\newcommand{\sig}{\ensuremath{\mathbb{\text{Sig}}}}
\newcommand{\sym}{\ensuremath{\mathbb{\text{Behav}}}}

\newcommand{\hexit}[1]{\texttt{0x#1}}
\newcommand{\hexitl}[2]{\truncate{#1}{\hexit{#2}}}

\newcommand{\instruction}[1]{{\textit{\texttt{\uppercase{#1}}}}}

\newcommand{\myth}[0]{\texttt{mythril}\xspace}
\newcommand{\SIN}{\ensuremath{sws_{0}}}
\newcommand{\SINA}{\ensuremath{sws_{1}}}
\newcommand{\CGE}{\ensuremath{c_{ge}}}

\newcommand{\lastBlock}{\num{5700000}}

\newcommand{\lastBlockDate}{30 May 2018\xspace}

\newcommand{\UniqueRuntimeCodeInstances}{\num{111882}\xspace}

\newcommand{\ContractInstances}{\num{6684316}\xspace}

\lstdefinelanguage{Solidity}{
  keywords={typeof, enum, new, external, true, false, catch, function, event, modifier, internal, private, sender, require, revert, return, null, catch, public, returns, switch, var, if, in, while, do, else, case, break, contract, throw, value, data, sig, gas, gasprice, origin, coinbase, difficulty, blockhash, timestamp, number, gaslimit, assert, require, revert, addmod, mulmod, selfdestruct, sha3, keccak256, ecrecover, sha256, ripemd160, dataCopy, pragma, solidity, constant, for, continue, ?, :, storage, memory},
  ndkeywords={class, export, implements, import, this, payable, view, pure, now, msg, block, wei, day, call, tx, send, delegatecall, callcode, address, uint, bytes2, bytes, fixedMxN, bytes32, void, bytesN, bool, bytes20, uint8, uint16, int16, uint248, int, uint256, int256, bytes4, uint32, int8, bytes1, byte, bytes32, mapping, string, struct},   
  basicstyle=\scriptsize\ttfamily,
  keywordstyle=\color{blue},
  ndkeywordstyle=\color{orange},
  commentstyle=\color{cyan},
  numberstyle=\color{gray},
  stringstyle=\color{purple},
  breakatwhitespace=false,         
  breaklines=true,                 
  captionpos=b,                    
  keepspaces=true,                 
  numbers=left,                    
  numbersep=5pt,                  
  showspaces=false,                
  showstringspaces=false,
  showtabs=false,                  
  tabsize=2,
  comment=[l]{//},
}

\lstdefinelanguage{EthereumBytecode}{
    keywords={PUSH1, CALLDATALOAD, PUSH4, EQ,PUSH1, EQ, JUMPI, JUMPDEST},
    ndkeywords={0x4, 0xa9059cbb, 0x20, 0x4},   
    basicstyle=\scriptsize\ttfamily,
    keywordstyle=\color{blue},
    ndkeywordstyle=\color{orange},
    commentstyle=\color{cyan},
    numberstyle=\color{black},
    stringstyle=\color{black},
    breakatwhitespace=false,         
    breaklines=true,                 
    captionpos=b,                    
    keepspaces=true,                 
    numbers=none,                    
    numbersep=5pt,                  
    showspaces=false,                
    showstringspaces=false,
    showtabs=false,                  
    tabsize=2,
    comment=[l]{//},
}

\lstdefinelanguage{JavaScript}{
  keywords={typeof, new, true, false, catch, function, return, null, catch, switch, var, if, in, while, do, else, case, break, console, =>},
  ndkeywords={class, export, boolean, throw, implements, import, this},
  basicstyle=\scriptsize,
  keywordstyle=\color{blue},
  commentstyle=\color{cyan},
  numberstyle=\color{gray},
  stringstyle=\color{purple},
  breakatwhitespace=false,         
  breaklines=true,                 
  captionpos=b,                    
  keepspaces=true,                 
  numbers=none,                    
  numbersep=5pt,                  
  showspaces=false,                
  showstringspaces=false,
  showtabs=false,                  
  tabsize=2,
  comment=[l]{//},
}

\usepackage{chngcntr}
\AtBeginDocument{\counterwithout{lstlisting}{chapter}}

\newcommand{\sol}[1]{{\lstinline[columns=fixed, language=Solidity]$#1$}}

\begin{document}

\title{Detecting Token Systems on Ethereum}


%


\author{Michael Fr"owis\inst{1} \and Andreas Fuchs\inst{2} \and Rainer B"ohme\inst{1,2}}

\institute{
Department of Computer Science, Universit\"at Innsbruck, Austria \\
\email{michael.froewis@uibk.ac.at}\and
Department of Information Systems, University of M\"unster, Germany
}
            
\maketitle

\begin{abstract}
We propose and compare two approaches to identify smart contracts as token systems by analyzing their public bytecode. 
The first approach symbolically executes the code in order to detect token systems by their characteristic behavior of updating internal accounts.
The second approach serves as a comparison base and exploits the common interface of \ERC, the most popular token standard. 
We present quantitative results for the \ET blockchain, and validate the effectiveness of both approaches using a set of curated token systems as ground truth. 
We observe 100\% recall for the second approach. Recall rates of 89\% (with well explainable missed detections) indicate that the first approach may also be able to identify ``hidden'' or undocumented token systems that intentionally do not implement the standard.
One possible application of the proposed methods is to facilitate regulators' tasks of monitoring and policing the use of token systems and their underlying platforms.
\end{abstract}


\section{Introduction}
\label{sec:intro}

Arguably, it has been easier to create a virtual asset on \ET in 2017 than a website on the Internet in 1997.
%
In September 2018, the market valuation of the well observable virtual assets (``tokens'') on the \EP amounts to US\$ 35 billion, not counting the US\$ 17.6 billion of ether, 
the platform's hardwired cryptocurrency.\footnote{Sources: Etherscan.io and Coinmarketcap.com on 12 September 2018, own calculations} These figures are the result of the 2017 boom of \emph{initial coin offerings} (ICOs), enabled by a combination of a hype around blockchain technology, lack of attractive conventional investment alternatives, 
and greed.

The sheer amount of money involved calls for regulators to take note and, where necessary, step in. While governments' concerns with cryptocurrencies, such as Bitcoin, were mainly focussed on tracking payment flows of criminal origin (\eg from trade with illegal goods, ransomware, money laundering, terrorism financing), the vast growth of an investment universe in virtual assets poses new challenges. These include enforcement of security laws \cite{SEC2017}, consumer protection \cite{Underwood2018}, and prudential monitoring in the interest of financial stability \cite{FSB2018}. 
These tasks require proven methods and adequate tools to detect, classify, and monitor virtual assets on platforms that can in principle host any kind of decentralized application. Therefore, in this work we set out to offer a scientific approach for the relevant case of token detection on \ET. 


In jargon, \emph{token} is a shorthand for a transferrable virtual good. The community 
distinguishes fungible from non-fungible tokens. Although the notion of fungibility is not precisely defined for all corner cases, a token is said to be fungible if all units are alike, i.\,e., each unit is interchangeable with every other unit.
By contrast, a non-fungible token has an identifying feature, such as a serial number, color, \ETC

Typical token systems on \ET are computer programs that allow its users to exchange tokens with each other in a decentralized, secure, and atomic way, up to the extent enforceable by the underlying blockchain-based system.
Such tokens can be useful in many scenarios.
For instance, fungible tokens can serve as means of payment~(e.\,g., sub-currencies), securitized rights (e.\,g., to vote or claim profit), or store of value.   
Non-fungible tokens are virtual collectibles.

Our approach is novel in that we detect fungible \TOKS by the characteristic program behavior, which is related to the secure exchange functionality. The behavior is detected by combining symbolic execution and taint analysis, two established static code analysis techniques, which were adapted to the application. As a comparison base, we also propose a signature-based  detection method that searches for instances of standard interfaces for \TOKS. We compare the effectiveness of both methods on a curated ground truth dataset before we generalize and present results for the entire \ET blockchain.

The paper is organized as follows. 
The next Section~\ref{sec:background} introduces necessary background. 
Sections~\ref{sec:method_behavior} and \ref{sec:method_signature} present our behavior-based and signature-based methods, respectively.
Performance measurements are reported and discussed in Section~\ref{sec:results}. 
Section~\ref{sec:related_work} connects to relevant related work, before Section~\ref{sec:conclusion} concludes with a discussion and an outlook to future applications and research directions.

\section{Background and Principles}
\label{sec:background}

This section recalls relevant properties of the \EP, specifically its virtual machine and calling conventions. It further sets up the static analysis techniques: symbolic execution, taint analysis, and the Ethereum call graph.


\subsection{Ethereum Virtual Machine (EVM)}
\label{sec:background_evm}

Ethereum is a decentralized system that updates a global state in a public, append-only data structure called \textit{blockchain} \cite{wood2017ethereum}.
At every point in time, the global state is an injective mapping from addresses to account states. 
Account states include the balance in ether, permanent storage, and optionally code controlling the account.
By convention, accounts with code are called \textit{smart contracts}, whereas accounts without code are called \textit{externally owned accounts}. 
\textit{Transactions} sent to the Ethereum network update the global state.
A transaction can (1) transfer ether between accounts, (2) create new accounts, 
(3) invoke code of any smart contract of the current state, or combinations thereof.
Arguments can be passed to code by supplying \textit{input data} in the transaction.

The Ethereum Virtual Machine (EVM) is a stack-based virtual machine that executes the bytecode in account states. 
Single-byte opcodes are followed by an optional immediate argument of length between 1 and 32 bytes. 
To prevent long-running or infinite computations, \ET charges a fee for every instruction executed, accounted in units of \emph{gas}.
Most developers program the \EVM in \emph{Solidity}, a high-level imperative programming language.

\subsection{Ethereum Application Binary Interface (ABI)}
\label{sec:background_abi}

The Application Binary Interface (ABI) specifies the calling conventions between \SCS.
Since the \EVM has no native concept of functions, every transaction sent to a contract starts the execution at the same entry point. 
Function-like behavior is implemented by a \emph{function dispatching} mechanism, which evaluates the leading 4 bytes of the input data. 
%
Specifically, every function is identified by a 4-byte \emph{function selector}, which is deterministically derived from the hash value of the \emph{function signature}.
A function signature is a concatenation of the function name and a list of argument types as defined in Solidity. 
For example, \texttt{transfer(address,uint256)} is a signature for a function called ``transfer'' accepting two arguments of type ``address'' and unsigned 32-byte integer, respectively.

Listing~\ref{lst:FunctionDispatch} illustrates the function dispatching mechanism in \EVM bytecode as generated by the \SOL compiler.
The full ABI definition can be found at~\cite{abidef}.

\begin{minipage}{\linewidth} 
\begin{lstlisting}[language=EthereumBytecode, caption={\EVM bytecode illustrating the ABI function dispatching.},label={lst:FunctionDispatch},escapechar=|]
  4 : PUSH1 0x4        // Push constant 4 on stack
  5 : CALLDATALOAD     // Load first 4 bytes from input data
  6 : PUSH4 0xa9059cbb // Function selector transfer(address,uint256)
  7 : EQ               // Check equality
  8 : PUSH1 0x20       // Push jump target 0x20 = 32
  9 : JUMPI            // Jump if true (cf. line 7)
  10: PUSH1 0x4        // If not equal, continue with this instruction
  ...
  32: JUMPDEST         // Implementation of transfer(address,uint256)
  33: ...
\end{lstlisting}
\end{minipage}

The ABI specification is not part of the \ET protocol. Anyone is free to define their own calling conventions. 
However, to our knowledge, all popular compilers targeting the \EVM produce ABI-compliant bytecode. 

\subsection{Symbolic Execution and Taint Analysis}
\label{sec:background_sym}

Symbolic execution is a program analysis technique \cite{king1976symbolic}. 
%
In contrast to concrete execution, symbolic execution does not only explore one execution path through a program by using concrete inputs, but tries to explore \emph{all} paths in a systematic manner. 
Program inputs 
are therefore represented as symbols.
The symbolic execution engine executes instructions akin the actual runtime environment as long as no symbolic values are involved. 
When an instruction depends on at least one symbolic value, the symbolic execution engine cannot execute the instruction directly, but builds a symbolic expression that describes the execution result. 

Special consideration is needed when it comes to control flow. 
Whenever a conditional branch is reached that depends on a symbolic branch condition $c$ within path $\pi_n$, the engine cannot decide which path to follow. 
Consequently, it follows both ($\pi_{n|true} \gets \pi_n \land c$, $\pi_{n|false} \gets \pi_n \land \overline{c}$) execution paths using backtracking.
To avoid the exploration of impossible paths, typical engines use an SMT solver to find a satisfying assignment for the path condition in question. 
If a suitable assignment is found the path is further explored.


For example, when the code in Listing~\ref{lst:FunctionDispatch} is symbolically executed with initial path constraint $\pi \gets true$,
the symbolic execution engine generates a path constraint $\pi_{true} \gets \delta$~$=$~\texttt{0xa9059cbb} 
for the path $\langle...,4,5,6,7,8,9,32,...\rangle$,
where $\delta$ is a symbolic variable representing the first four bytes of the input data.
When the symbolic execution of the path corresponding to $\pi_{true}$ completes, 
the symbolic execution engine performs backtracking, 
generates a constraint $\pi_{false} \gets \delta$~$\neq$~\texttt{0xa9059cbb},
and continues on the path $\langle...,4,5,6,7,8,9,10,...\rangle$.

In this work we mainly exploit two properties of symbolic execution.
First, we use the explored paths as input to \emph{static taint analysis}\cite{schwartz2010all}. 
Taint analysis is a technique to trace data flows of interest through a program execution.
More concretely, we label user inputs with markers (``taint'') and track which storage locations are affected by it.
Our second use of symbolic execution is to access the structure of symbolic expressions generated by the engine.


Symbolic execution faces many limitations in practice \cite{baldoni2018survey}: path explosion, unbounded loops, and the NP-hardness of the SMT problem all require tradeoffs, such as imposing timeouts and skipping paths. The success of symbolic execution can be measured in terms of code coverage. 
Gladly, most smart contracts on \ET are very short programs, gas makes unbounded loops expensive, and therefore \ET is more amenable to symbolic execution than other platforms. 

 



\subsection{Ethereum Call Graph}
\label{sec:background_ecg}

Both detection methods introduced in this work operate locally.
This means we only analyze code of one address at a time. 
Consequently, the methods are blind to behavior or signatures located outside the smart contract under analysis. Recall from Section~\ref{sec:background_evm} that transactions can invoke code of any smart contract active in the current state. 
Smart contracts can create transactions using the call family\footnote{\instruction{CALL}, \instruction{DELEGATECALL}, \instruction{CALLCODE}, \instruction{STATICCALL}} of instructions.
Such calls are used in smart contracts to 
(1) interact with other parties (smart contracts), and 
(2) reuse code already deployed.

A useful tool to look beyond the local address is the \ET call graph~\cite{frowis2017code}.
It holds information on relationships between contracts obtained by parsing all bytecode on the \ET blockchain and extracting all statically encoded addresses used in instructions of the call family. 
Nodes in the graph are addresses with code. Directed edges denote static calls from caller to callee.

The so-constructed call graph captures only statically encoded references. References to other contracts set on construction, calculated at runtime, or provided as user input are missed. The only practical way to work around this limitation is dynamic analysis, which makes a different trade-off as it limits the analysis to actually executed rather than all possible paths.


\section{Behavior-based Token Detection}
\label{sec:method_behavior}
Now we describe our behavior-based heuristic detection method for fungible token systems on the \EP. We first justify the behavioral pattern, then present our detection method, and finally discuss known limitations.

\subsection{Pattern}
\label{sec:method_behavior_pattern}
Fungiblity means that all tokens in a given token system are alike.
As a result, token systems do not need to store which specific token belongs to which party. 
The only relevant information is who owns how many tokens.
A straightforward (and gas-efficient) way to implement the state of a token system is storing a mapping of owners (identified by addresses) to a non-negative number of tokens. 

An important property of \TOKS is the ability to transfer tokens. 
We assume that a token system wants to preserve the total amount of tokens in circulation as they are transferred. 
%
In order to detect \SCS that behave like \TOKS we define:
\begin{definition} A \emph{token system} according to its behavior, is a smart contract that
(1) stores users' balances as integers in permanent storage, and
(2) provides a function to transfer tokens between users while keeping the total balance constant, where 
(3) the transferred value is controlled by user input.
\label{def:TOK}
\end{definition}


Fixing the data type to integers in (1) is reasonable as the \EVM does not natively support floating point or rational numbers.

\begin{minipage}{\linewidth} 
\begin{lstlisting}[language=Solidity, caption={Transfer pattern in \SOL, typical for fungible tokens.},label={lst:TokenPattern},escapechar=|]
contract FungibleTokenPattern {
  mapping(address => uint) balance; |\label{line:balance}|

  function sendToken(address to, uint value) public {
      require(balance[msg.sender] >= value);  |\label{line:constraint}|
      balance[msg.sender] = balance[msg.sender] - value;           |\label{line:sender}|
      balance[to] = balance[to] + value;                   |\label{line:rec}|
  }
}
\end{lstlisting}
\end{minipage}

Listing~\ref{lst:TokenPattern} shows a \SOL implementation of a minimalistic token system that complies with Definition~\ref{def:TOK}.

\subsection{Detection Method}
\label{sec:method_behavior_method}

We propose an approach that analyzes the behavior of potential token systems based on symbolic execution and static taint analysis. 

Our approach works as follows. 
We look for a possible execution path 
that updates two integers in storage, one for the sender and recipient, by a value defined as parameter.
For (1) and (3) of Definition~\ref{def:TOK}, we use taint analysis to find storage write states ($sws$), where the value stored can be influenced by user input. 
For each of those stores of input data \SIN\xspace, we try to find a matching store \SINA\xspace that follows our tainted store on some possible execution path.
Furthermore, we look for constraints in the path condition that check if the value in some storage field is larger than or equal to some user input field (\CGE).
This captures the check that the sender's balance cannot be negative.
We organize our stores and path constraint in triplets of the form $(\SIN, \SINA, \CGE)$, meaning we found a storage write \SIN\xspace with its value influenced by user input. 
\SIN\xspace is followed by \SINA\xspace on some viable execution path.
Additionally, we have a constraint \CGE\xspace on this path that checks if some storage field is larger than a user input field.
We call such a triplet \emph{transfer candidate}.
What remains to verify is (2), i.\,e., whether the operations on \SIN\xspace and \SINA\xspace are really transferring value and if \CGE\xspace is a constraint on one of the fields written to.
Here we apply a heuristic that looks at the term structure of transfer candidates.

Algorithm~\ref{alg:symbolic_triplets} shows the analysis done for every possible transfer candidate triplet.
We use $\vartriangleleft$ and $\trianglelefteq$  to denote the proper subterm and subterm relation. 
\begin{algorithm}[ht]
  \caption{Analyzing transfer candidates.}
  \label{alg:symbolic_triplets}
  \begin{algorithmic}
\Function{IsToken$_{sym}$}{$S$} \Comment{a set $S$ of triplets $(sws_{0},sws_{1}, c_{ge})$}
  \State $s_{ops} \gets \{+, -\}$

  \For{$(sws_{0},\,sws_{1},\,c_{ge}) \in S$}
    \State $b_{rss},\,s_{opsLeft},\,b_{usedC} \gets$ \Call{CheckStoreTerm}{$sws_{0},\,s_{ops}, true$}
    \If {$b_{rss} \land |s_{opsLeft}| = |s_{ops}| - 1$}
      \State $b_{rss},s_{opsLeft},b_{usedC} \gets$ \Call{CheckStoreTerm}{$sws_{1},\, s_{opsLeft}, \neg b_{usedC}$}
      \If {$b_{rss} \land s_{opsLeft} = \emptyset \land b_{useC}$}
        \State \Return $true$ \hfill\Comment{Found a token-like behavior.}
      \EndIf
    \EndIf
  \EndFor

  \State \Return $false$ \hfill\Comment{None of the candidates indicates a tokens system.}
\EndFunction
\Function{CheckStoreTerm}{$sws_{n}$,\,$c_{ge}$,\,$s_{ops}$, $b_{cToEqC}$} 

  \State $b_{selfRef},\,b_{callData},\,b_{toEqC} \gets   false$
  \State $t_{to},\,t_{val}\,\,\,\gets   sws_{n}.to,sws_{n}.value$ \Comment{Store has an address and a value.}
  \State $s_{opFirst}\,\,\gets \Call{FindFirstOpBFS}{t_{val}, s_{ops}}$         \Comment{Get first matching function symbol.}
  \State $b_{selfRef}\,\,\gets  t_{to} \vartriangleleft t_{val}$\hfill\Comment{Store updates itself?}
  \State $b_{callData}\gets  c_{ge}.smallerTerm \vartriangleleft t_{val}$\hfill\Comment{Term contains input from constraint?}
  \State $b_{toEqC}\,\,\,\,\,\gets t_{to} \trianglelefteq c_{ge}.largerTerm$\hfill\Comment{Is constraint on assignment?}

  \State \Return $(t_{selfRef} \land t_{calldata},\, s_{ops} \setminus s_{opFirst},\, (b_{cToEqC} \land b_{toEqC}) \lor \neg b_{cToEqC})$
\EndFunction
  \end{algorithmic}
\end{algorithm}

\textbf{Example:}
We use the example contract in Listing~\ref{lst:TokenPattern} to illustrate how the algorithm works.
We refer to source code when possible, although the actual analysis is done on bytecode.
First we perform taint analysis to find storage writes influenced by user input.
We find stores in lines \ref{line:sender} and \ref{line:rec}. 
Then we look for followup stores along a viable execution path. 
Only for the store in line \ref{line:sender} we find a following store, namely in line~\ref{line:rec}.
Furthermore, we look at path conditions at the program state of the first store in line~\ref{line:sender}. 
We find one suitable condition that matches our restrictions that the condition checks if a storage field is larger than (or equal to) some user input in line~\ref{line:constraint}.
This means we found one transfer candidate to check $(line~\ref{line:sender}, line~\ref{line:rec}, line~\ref{line:constraint})$.
First we execute \textsc{CheckStoreTerm} on \sol{balance[msg.sender] = balance[msg.sender] - value} with \sol{balance[msg.sender] >= value} as a constraint and $\{+,-\}$ as possible operations, and $b_{cToEqC} = true$ .
Then we check if the \emph{right hand side} (RHS) of the store term contains itself, which it does.
It follows a check if the RHS of the constraint \sol{value} is a subterm of our store term, meaning that the constraint and store refer to the same user input.
If that is the case, we check if we can either find an addition or subtraction in our term. 
\textsc{FindFirstOpBFS} checks all function applications in the term against a list of operations (starting with $\{+,-\}$) and returns a set with the first operation to occur, or an empty set if the operations are not found. 
Finally, we check if the storage field used in the constraint is the target of the store, which is true in our case.
The function then returns a tuple with $(true, \{+\}, true)$, since the terms of our transfer candidate fulfill all conditions. We found that minus is the root operation on the term and already found the constraint value to be written on.
We then continue with calling \textsc{CheckStoreTerm} again for the second term, with a reduced list of operations, only looking for plus and no longer looking for writes on our constraint values.
This call returns $(true, \emptyset, true)$, thus we found token-like behavior according to our definition.

\subsection{Known Limitations}
\label{sec:method_behavior_example}

We inherit the limitations from symbolic execution (cf.~Sect.~\ref{sec:background_sym}).
%
We use \myth \cite{mythril}, 
a tool designed for security analyses that is known to reach high accuracy \cite{parizi2018empirical} despite using heuristics.
For our experiments, we run \myth with a timeout of 60 seconds and a maximum path length of 58. 
 %
Furthermore, mythril is under active development and has a couple of limitations that may influence our results and their replicability.
For example in taint analysis,
the current version of \myth (0.18.11) cannot spread taint over storage or memory fields. 
This can cause problems when function parameters are passed by reference. 

The locality is dealt with in the following way: whenever the symbolic execution reaches a call, we consider it as communication with the unknown environment. Hence, the engine introduces a fresh unrestricted symbol for the return value and carries on. 
That means the analysis is blind to everything that happens outside of the code of the current address. 
We evaluate the impact of this limitation empirically with the call graph in Section~\ref{sec:results_all}.

%
Another limitation lies in the definition of the pattern.
It is not straightforward to find the best approximation for the behavior we search for, since the same behavior can be implemented in various ways that may result in vastly different bytecode.
What eases this problem somewhat is that much of the bytecode currently deployed on \ET is produced by a pretty homogeneous toolchain (\SOL and \texttt{solc}). 
Moreover, gas favors simple programs, often rendering abstractions that would complicate the underlying bytecode uneconomic.



\section{Signature-based Token Detection}
\label{sec:method_signature}
Now we present a simple signature-based heuristic to detect \TOKS. It evaluates if the bytecode implements the ABI standard for the \ERC interface.
We need this method as a benchmark to evaluate the behavior-based approach. 

\subsection{Pattern}
\label{sec:method_signature_pattern}

To improve the interoperability of tokens in the \ET ecosystem, the community has established a set of standards for \TOKS.
\ERC~\cite{ercstd} is the most popular standard for fungible tokens. It also serves as basis for extensions, such as ERC-223 and ERC-621.
Even ERC-777, while still at draft stage at the time of writing, is backward compatible: a token can implement both standards to interact with older systems that require the \ERC interface \cite{ERC777}. Given the vast dominance of \ERC today, we restrict our analysis to this standard.


The standard defines six functions and two events that must be implemented to be fully compliant. (Listing~\ref{lst:ERC20} in Appendix~\ref{app:erc} shows the \ERC interface skeleton in \SOL.)
Since the applications of tradable tokens are diverse, the standard does not define how tokens are created, initially distributed, or how data storage should be organized.
It only defines that \ERC tokens must have functions to securely transfer tokens, and some helper functions to check balances.

\subsection{Detection Method}
\label{sec:method_signature_strategy}

A na\"ive way to detect tokens is to check if the code implements the methods defined by the \ERC standard.
From the ABI definition~(see Sect.~\ref{sec:background_abi}) we know how function calls are encoded and how functions are dispatched.
%
In order to detect \TOKS based on a signature we define:
\begin{definition} A \emph{token system} according to its signature is a smart contract that
introduces at least 5 of the 6 function selectors defined by the \ERC standard.
\label{def:TOK2}
\end{definition}
We used five as a threshold to account for incomplete implementations of \ERC. 

We use Definition~\ref{def:TOK2} and the fact that the only way to introduce constants in the \EVM are \instruction{PUSH} instructions.
Since function selectors are 4 bytes long according to the ABI, the detection method looks for \instruction{PUSH4} instructions.
Algorithm~\ref{alg:signature} takes as input a list of \EVM instructions, inspects all 4-byte constants introduced, and checks membership in the pre-determined set of \ERC function selectors~(variable $s_{signatures}$).


\begin{algorithm}
\caption{Detection method based on disassembly and signatures.}
\label{alg:signature}
\begin{algorithmic}

\State $s_{signatures}\gets$ 
 $\{
    18160ddd, 
    70a08231, 
    dd62ed3e,
    a9059cbb, 
    095ea7b3, 
    23b872dd
\}$
\par
\Function{IsToken$_{\sig}$}{$I$} \Comment{$I$ is a  list of instruction tuples $t \in (opcode \times arg)$}
  \State $s_{constants}\gets \emptyset$
  \For{$(i_{opcode},\,i_{argument}) \in I$} 
    \If {$i_{opcode} = PUSH4$}
      \State $s_{constants} \gets s_{constants} \cup \{i_{argument}\}$
    \EndIf
  \EndFor \par

  \State \Return $|s_{constants} \cap s_{signatures}| \geq 5$
\EndFunction
\end{algorithmic}
\end{algorithm}

\subsection{Known Limitations}
\label{sec:method_signature_example}

This method is obviously prone to false positives if a contract pushes all required constants to the stack but never uses them. 
This may even happen in dead code. 
Hence, we also get false positives if we analyze so-called \emph{factory contracts} that create new token systems when called \cite{factroy1}.
The code of the factory includes the code of the token system to create, and thus contains push instructions of the required constants.\footnote{One such instance can be found at \hexit{bf209cd9f641363931f65c0e8ef44c79ca379301}}
The common cause for these weaknesses is that the method considers neither data nor control flow.

Similar to the behavior-based method, the signature-based method is a local heuristic.
This can result in false negatives. For example, if the \SC does not implement the \ERC interface, 
but delegates
calls to a suitable implementation.
This form of delegation is common practice on the \EP because it makes deployments cheaper. 
Furthermore, it enables code updates by swapping the reference to the actual implementation\cite{proxy1,proxy2}.



\section{Measurements}
\label{sec:results}


\subsection{Data and Procedure}
\label{sec:results_dataset}

To evaluate our two detection methods we study the \ET main chain from the day of its inception until \lastBlockDate.\footnote{Block number: \lastBlock}
We extract all \emph{unique runtime bytecode instances} 
and the addresses they are deployed on.
With \emph{runtime bytecode} we denote code that is executed when a transaction is sent to the contract after its deployment. This means we do not analyze initialization code.

In total we found \ContractInstances addresses that hosted bytecode at one point in time. From these addresses we extract \UniqueRuntimeCodeInstances \emph{unique runtime bytecode instances}, henceforth referred to as \emph{bytecode instances} for brevity, unless stated otherwise.
Observe that we do not double-count bytecode instances unlike it is often the case in headline statistics on \SCS. 
We do not exclude contracts that were disabled by selfdestruct, i.\,e., we analyze all code ever deployed.  
Consequently, we also analyze bytecode instances that are barely used. 

To evaluate that our detection results are not biased towards barely used or test code, we also define a subset of \emph{active} instances. 
We define a bytecode instance as \emph{active} if all hosting addresses combined handled a volume of at least 1000 transactions until \lastBlockDate.


To build a \emph{ground truth dataset}~(GTD) for the evaluation, we downloaded \num{612} Top \ERC tokens\footnote{Ranked by market cap, retrieved on 23 Aug. 18 from \url{https://etherscan.io/tokens}} from Etherscan.
Etherscan, a popular \ET block explorer, curates its top list of \ERC tokens by only including systems that are popular, supported by at least one major exchange, compliant with \ERC, and have a visible website.
We exclude all \TOKS that were created after \lastBlockDate, leaving us with a curated list of \num{595} ground truth \TOKS, of which we extract \num{578} bytecode instances. 

\newcommand{\SymAndSig}[0]{\ensuremath{\sym \land \sig}}
\newcommand{\SymOrSig}[0]{\ensuremath{\sym \lor \sig}}
\newcommand{\nSymAndSig}[0]{\ensuremath{ \overline{\sym} \land \sig}}
\newcommand{\SymAndnSig}[0]{\ensuremath{ \sym \land \overline{\sig}}}
\newcommand{\nSymAndnSig}[0]{\ensuremath{ \overline{\sym} \land \overline{\sig}}}
\newcommand{\bytecodeSizeAll}[0]{all instances \num{3315.0} (\num{2541})}

We run both of detection methods over all bytecode instances and evaluate the results.
The signature-based method~(\sig) is able to process all of the input contracts. 
The behavior-based method~(\sym) fails to analyze \num{1373}~(\num{1.23}\% 
of the total) instances.
Failures occur if, for example, the bytecode contains syntactic errors not handled in the engine. 
We consider those \num{1373} instances as negative detection results. On the remaining \num{110509} 
instances, our behavior-based method reaches a mean (median) code coverage of  \num{71,9}\% (\num{82,2}\%). 
Over \num{70}\% of the instances reach a coverage above \num{50}\%, supporting the claim that smart contracts are a very suitable for symbolic execution techniques.\footnote{100\% - \#\num{16056}, $\leq 75\%$ - \#\num{50874}, $\leq 50\%$ - \#\num{31298}, $\leq 25\%$ - \#\num{5165}, $\leq 10\%$ - \#\num{1031}} 

To confirm our restriction to the \ERC interface in the signature-based method, we adapted our method to count the number of ERC-777 tokens. 
We encounter only four systems implementing at least 4 of the 13 functions required by ERC-777. All of them also implement \ERC for backward compatibility. 


\subsection{Validation on Ground Truth Data and Error Analysis}
\label{sec:results_detection}

\newcolumntype{C}{>{\centering\arraybackslash} m{3cm}}

\newcommand{\DetectTable}[9]{
  \begin{tabular}{ l | C | C | r }
    &  \multicolumn{1}{p{3cm}|}{\centering \textbf{Detected} \\ \textbf{by \sig}} & \multicolumn{1}{p{3cm}|}{\centering \textbf{Not Detected} \\ \textbf{by \sig}} & \\
    \hline
    \multicolumn{1}{p{3cm}|}{\centering \textbf{Detected} \\ \textbf{by \sym}} & \multirow{ 2}{*}{#1}  & \multirow{ 2}{*}{#2}  & \cellcolor{lightgray!60}\multirow{ 2}{*}{{#3}} \\
    \hline
    \multicolumn{1}{p{3cm}|}{\centering \textbf{Not Detected} \\ \textbf{by \sym}} &  \multirow{ 2}{*}{#4} & \multirow{ 2}{*}{#5} &\cellcolor{lightgray!60} \multirow{ 2}{*}{{#6}}\\
    \hline
    \multicolumn{1}{p{3cm}|}{}  & \cellcolor{lightgray!60} \multirow{ 2}{*}{{#7}} & \cellcolor{lightgray!60} \multirow{ 2}{*}{{#8}} & \cellcolor{lightgray!60} \phantom{as} \multirow{ 2}{*}{{#9}} \\[10pt]
  \end{tabular}
}


\begin{table}[t]
  \centering
  \caption{Recall of signature- and behavior-based detection methods against our GTD.}
  \label{tab:detection_gt}
  \DetectTable{\num{87.89}\% (508)}{\num{0.87}\% (5)}{\num{88.75}\% (513)}
              {\num{11.25}\% (65)}{\num{0.00}\% (0)}{\num{11.25}\% (65)}
              {\num{99.13}\% (573)}{\num{0.87}\% (5)}{\num{100.00}\% (578)}
\end{table}

Table~\ref{tab:detection_gt} presents the detection results of both methods evaluated against our curated GTD.
Observe that the signature-based method alone is pretty good at detecting tokens, reaching \num{99.13}\% recall. The behavior-based method performs visibly worse with a recall of \num{88.75}\% 
on our curated GTD. Since no \TOKS remained undetected by both methods, the combination of both (\SymOrSig) gives us perfect \num{100}\% recall. 
Our GTD does not allow us to calculate the precision.

The behavior-based method is able to detect the exact five contracts that are missed by the signature-based approach (\SymAndnSig). 
Further manual investigation of these contracts shows that all of them do not implement \ERC up to our threshold. 
Fortunately, all of the five contracts published \SOL source code.
Thus, we could confirm that they are missed by the signature-method because they implement only three of the six \ERC functions, namely \sol{totalSupply()}, \sol{balanceOf(address)}, and \sol{transfer(address,uint256)}.
This suggests that our initial threshold is too high, or in other words that even major token systems handle standards laxer than expected. 
Table~\ref{tab:gt_sym_but_not_sig} in Appendix~\ref{app:manual} lists those five contracts.


The signature-based method identified \num{65} 
\TOKS that were not found by the behavior-based method (\nSymAndSig).
We conjecture that either those tokens implement their internal state differently or they use libraries that implement the bookkeeping of storage values, thereby escaping our local behavior-based analysis.
We try to answer why those tokens are not detected by manually inspecting a random sample of 20 (out of \num{65}) 
bytecode instances (listed in Table~\ref{tab:gt_sig_but_not_sym} in Appendix~\ref{app:manual}). 
We find that all of them are large and reasonably complex contracts.\footnote{Mean (median) code size: \bytecodeSizeAll, \nSymAndSig\xspace \num{8153.86} (\num{7828}) \\ Code coverage: \num{41.75}\% (\num{40.25}\%)} 
We encountered three main causes for missed detection:

\textbf{Delegation of Bookkeeping (6)}: We found 6 bytecode instances in our sample that do not implement any asset management logic in the contract itself. 
It is delegated to another contract.
The front-end contract implements the \ERC interface, but many back-end bookkeeping contracts do not, \eg the \textit{Digix Gold Token}. 
Delegation patterns (or ``hooks'') like this one are often used to allow updates (by reference substitution) of the asset management logic.

\textbf{Violation of Definition~\ref{def:TOK} (10)}: The second reason concerns mainly tokens that are derived from the popular MiniMeToken~\cite{minime}.
We found 9 of those in our sample. 
MiniMe uses a different storage layout.
Instead of a plain integer that is updated over time, it writes a new checkpoint for every transfer into an array. 
This violates our detection assumption (1) in Definition~\ref{def:TOK}, or, more specifically, fails our check that the field gets updated (self reference).
Even though this already defeats our detection method, we find that \myth was not able to inspect the relevant paths in the transfer function.
The average code coverage is as low as \num{34.8}\% on the 9 MiniMe-based tokens in our sample.

Also the \textit{MakerDAO} instance is not detected for a violation of Definition~\ref{def:TOK}, although it is not derived from MiniMe and reached high coverage (\num{94.3}\%).
It does not implement a balance check before the actual transfer as required in Definition~\ref{def:TOK}. This can be fixed with an ad-hoc adjustment of the method, but we are concerned about the (not observable) false positive risk of a relaxed behavioral pattern.


\textbf{Litations of Symbolic Execution and Taint Analysis (4)}: Four contracts in our sample neither delegate the bookkeeping work nor are derived from the MiniMeToken.
All of them use a simple integer value to store the balance of the participants. \textit{Storiqua} (\num{42.9}\%),  \textit{LocalCoinSwap} (\num{34.82}\%), and  \textit{LOCIcoin} (\num{18,5}\%) suffer from low coverage.
In the \textit{Storiqua} instance, our method finds the relevant paths in the transfer function but does not find a matching store.
In the case of \textit{LocalCoinSwap}, we do not find a suitable constraint although the symbolic execution engine explores the relevant paths in transfer.
\textit{LOCIcoin} has the lowest coverage. The engine does not discover the relevant paths in the transfer method.
Finally, \textit{TrueUSD} reaches high coverage (\num{72.9}\%), but the behavior-based method did not find a suitable constraint in the transfer function.
All of those cases are examples for known limitations of symbolic execution (reaching low coverage, missing relevant paths), taint analysis (failing to find matching stores), as well as our detection approach (missed constraints).

\subsection{Generalization to All Smart Contracts on Ethereum}
\label{sec:results_all}

Table~\ref{tab:detection} reports detection statistics over all bytecode instances. 
We find that \num{33.17}\% 
of the bytecode instances on the \EP can be said to be \TOKS with high confidence because they are detected by both methods ($\sig \land \sym$). The interesting part is where both methods disagree. 

Recall from our manual ground truth analysis that all instances missed by the signature-based method but detected by the behavior-based method (\SymAndnSig) are caused by our high threshold. So we re-run the analysis with a lower threshold of 3, as our manual inspection suggested. 
Table~\ref{tab:detection_gt_tresh} and Table~\ref{tab:detection_thresh} (both in Appendix~\ref{app:tabs}) show the updated results of Table~\ref{tab:detection_gt} and Table~\ref{tab:detection}, respectively.
With the lower threshold, the signature-based method detects \num{7193} more bytecode instances as tokens. 
\num{3232} of those newly detected token systems were already identified by the behavior-based method. The remaining tokens would have been missed otherwise. 
We conjecture that the \num{1772} bytecode instances only detected by the behavior-based method (\SymAndnSig)  
are either non-\ERC bookkeeping contracts, as found in the \textit{Digix Gold Token}, or \TOKS that do not implement \ERC for other reasons, such as obscuring their nature.


In the case of token systems detected by the signature-based but not by the behavior-based method (\nSymAndSig), we found mixed reasons in our GTD.
First, we saw systems that implement the \ERC interface but delegate all bookkeeping tasks to other contracts.
In order to study if this pattern generalizes to the whole dataset, we extract bytecode metrics, such as the number of call instructions.
We find that contracts that are detected by the signature-based method contain an above-average number of call instructions.
Table~\ref{tab:avg_calls} (in Appendix~\ref{app:tabs}) presents the mean and median values of call-family instructions for different subsets of bytecode instances. 
The highlighted row stands out: \nSymAndSig\xspace instances have on average \num{2.2} times as many calls as the average bytecode instance. 
This indicates the use of delegation patterns as found in the \textit{Digix Gold Token}. 
To further strengthen this interpretation, we us the \ET call graph (cf.~Sect.~\ref{sec:background_ecg}) to find out if those instances have calls to other instances that are otherwise classified as token systems. 
For \num{920}  of \num{10472} instances (\nSymAndSig) we find static references. \num{563} have direct hardcoded calls to another instance classified as token system, suggesting that the detectable behavior is implemented in the callee. 
The second and third reason for missing tokens were inherent limitations of symbolic execution, which we could not further evaluate on the entire dataset.

Table~\ref{tab:detection_thresh_tx} shows detection results for all \emph{active} bytecode instances.
The results are pretty comparable to Table~\ref{tab:detection_thresh}.
Note that the behavior-based method misses relatively more instances detected via signature than on the complete dataset.
One interpretation is that high-profile tokens implement more complex logic, therefore evading detection by symbolic execution. 
This conjecture is supported by the observed bytecode sizes as well as code coverage reached: active bytecode instances are on average around \num{1.6} times as large as the average bytecode instance. Average code coverage also drops from \num{71.9} (\num{82.2}\%) to \num{66.2} (\num{69.6}\%). 
\begin{table}[t]
  \centering
  \caption{Comparison of signature- and behavior-based detection methods on \emph{all} bytecode instances. (Note that this is not a performance measurement. We cannot expect 100\%.)}
  \label{tab:detection}
  \DetectTable{\num{33.17}\% (\num{37114})}{\num{4.47}\%  (\num{5004})}{\num{37.65}\% (\num{42118})}
              {\num{5.82}\% (\num{6512})}{\num{56.53}\% (\num{63252})}{\num{62.35}\% (\num{69764})}
              {\num{38.99}\% (\num{43626})}{\num{61.01}\% (\num{68256})}{\num{100.00}\% (\num{111882})}
\end{table}

\begin{table}[t]
  \centering
  \caption{Comparison of signature- and behavior-based detection methods on all \emph{active} bytecode instances (with signature threshold $\ge 3$).}
  \label{tab:detection_thresh_tx}
  \DetectTable{\num{32.56}\% (\num{2052})}{\num{0.73}\% (\num{46})}{\num{33.29}\% (\num{2098})}
              {\num{18.15}\% (\num{1144})}{\num{48.56}\% (\num{3060})}{\num{48.56}\% (\num{4204})}
              {\num{50.71}\% (\num{3196})}{\num{49.29}\% (\num{3106})}{\num{100.00}\% (\num{6302})}
\end{table}



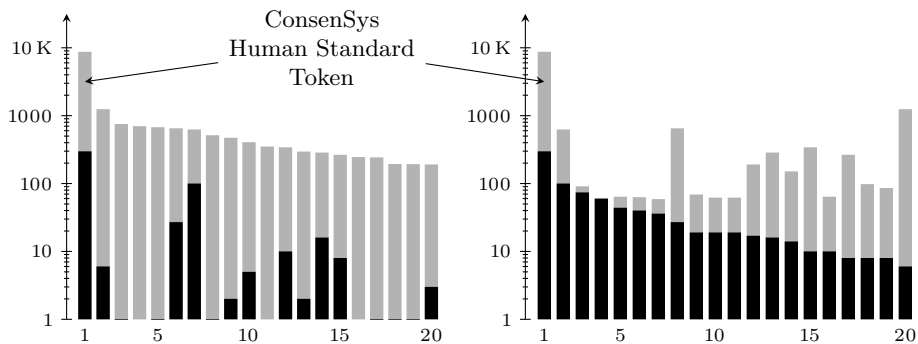
\begin{figure}
\begin{center}
\begin{tikzpicture}[>=stealth,x=2.4mm,y=9mm,remember picture]


    \draw [->] (0,0)--(0,4.5);
    \foreach \i in {1,5,10,15,20}
        \draw (\i,0) node [below] {\scriptsize \i};


    \draw (2pt,4)--++(-4pt,0) node [left] {\scriptsize 10\,K};
    \foreach \y/\l in {0/1,1/10,2/100,3/1000}
    {
        \draw (2pt,\y)--++(-4pt,0) node [left] {\scriptsize \l};
        \foreach \k in {0.301,0.4771,0.6021,0.699,0.7782,0.8451,0.9031,0.9542} 
            \draw (0,\y)++(1pt,\k)--++(-2pt,0);
    }
    

    \begin{scope}[line width=1.75mm]
        \clip (0,0) rectangle (21,4.5); 
        \draw [black!30] (1,0)--++(0,3.9409644934928) 
        (2,0)--++(0,3.09551804232315) 
        (3,0)--++(0,2.87621784059164) 
        (4,0)--++(0,2.84509804001426) 
        (5,0)--++(0,2.82994669594164) 
        (6,0)--++(0,2.81358098856819) 
        (7,0)--++(0,2.79657433321043) 
        (8,0)--++(0,2.71180722904119) 
        (9,0)--++(0,2.67394199863409) 
        (10,0)--++(0,2.60959440922522) 
        (11,0)--++(0,2.54530711646582) 
        (12,0)--++(0,2.53402610605613) 
        (13,0)--++(0,2.47129171105894) 
        (14,0)--++(0,2.45636603312904) 
        (15,0)--++(0,2.42324587393681) 
        (16,0)--++(0,2.39093510710338) 
        (17,0)--++(0,2.38560627359831) 
        (18,0)--++(0,2.28780172993023) 
        (19,0)--++(0,2.28555730900777) 
        (20,0)--++(0,2.28103336724773) 
        ;

         \draw [] (1,0)--++(0,2.47421772143784)
        (2,0)--++(0,0.778223626766096)
        (3,0)--++(0,-3)
        (4,0)--++(0,0.000434077479318593)
        (5,0)--++(0,-3)
        (6,0)--++(0,1.43137984884194)
        (7,0)--++(0,2.0000043429231)
        (8,0)--++(0,-3)
        (9,0)--++(0,0.301247088636211)
        (10,0)--++(0,0.699056854547668)
        (11,0)--++(0,0.000434077479318593)
        (12,0)--++(0,1.00004342727686)
        (13,0)--++(0,0.301247088636211)
        (14,0)--++(0,1.20414712521285)
        (15,0)--++(0,0.903144270409538)
        (16,0)--++(0,0.000434077479318593)
        (17,0)--++(0,-3)
        (18,0)--++(0,-3)
        (19,0)--++(0,-3)
        (20,0)--++(0,0.477265995424853) 
        ;

    \end{scope}
    
    
    \draw (14,4) node (L) {\parbox{25mm}{\centering ConsenSys Human Standard Token}};
    \draw [->] (L) -- (1,3.5);
    
\end{tikzpicture}
\begin{tikzpicture}[>=stealth,x=2.5mm,y=9mm,remember picture]


    \draw [->] (0,0)--(0,4.5);
    \foreach \i in {1,5,10,15,20}
        \draw (\i,0) node [below] {\scriptsize \i};


    \draw (2pt,4)--++(-4pt,0) node [left] {\scriptsize 10\,K};
    \foreach \y/\l in {0/1,1/10,2/100,3/1000}
    {
        \draw (2pt,\y)--++(-4pt,0) node [left] {\scriptsize \l};
        \foreach \k in {0.301,0.4771,0.6021,0.699,0.7782,0.8451,0.9031,0.9542} 
            \draw (0,\y)++(1pt,\k)--++(-2pt,0);
    }
    

    \begin{scope}[line width=1.75mm]
        \clip (0,0) rectangle (21,4.5); 
        \draw [black!30] (1,0)--++(0,3.9409644934928) 
        (2,0)--++(0,2.79657433321043) 
        (3,0)--++(0,1.95904139232109) 
        (4,0)--++(0,1.77815125038364) 
        (5,0)--++(0,1.80617997398389) 
        (6,0)--++(0,1.79934054945358) 
        (7,0)--++(0,1.77085201164214) 
        (8,0)--++(0,2.81358098856819) 
        (9,0)--++(0,1.83884909073726) 
        (10,0)--++(0,1.79239168949825) 
        (11,0)--++(0,1.79239168949825) 
        (12,0)--++(0,2.28103336724773) 
        (13,0)--++(0,2.45636603312904) 
        (14,0)--++(0,2.17897694729317) 
        (15,0)--++(0,2.53402610605613) 
        (16,0)--++(0,1.80617997398389) 
        (17,0)--++(0,2.42324587393681) 
        (18,0)--++(0,1.99122607569249) 
        (19,0)--++(0,1.93449845124357) 
        (20,0)--++(0,3.09551804232315) 
        ;

         \draw [] (1,0)--++(0,2.47421626407626)
        (2,0)--++(0,2)
        (3,0)--++(0,1.86923171973098)
        (4,0)--++(0,1.77815125038364)
        (5,0)--++(0,1.64345267648619)
        (6,0)--++(0,1.60205999132796)
        (7,0)--++(0,1.55630250076729)
        (8,0)--++(0,1.43136376415899)
        (9,0)--++(0,1.27875360095283)
        (10,0)--++(0,1.27875360095283)
        (11,0)--++(0,1.27875360095283)
        (12,0)--++(0,1.23044892137827)
        (13,0)--++(0,1.20411998265592)
        (14,0)--++(0,1.14612803567824)
        (15,0)--++(0,1)
        (16,0)--++(0,1)
        (17,0)--++(0,0.903089986991944)
        (18,0)--++(0,0.903089986991944)
        (19,0)--++(0,0.903089986991944)
        (20,0)--++(0,0.778151250383644) 
        ;

    \end{scope}
    
        
    \draw [->,overlay] (L) -- (1,3.5);
    
\end{tikzpicture}
\end{center}
\caption{Bytecode reuse of \ET token systems: number of addresses hosting a unique bytecode instance detected as token system. Top-20 ranked by total addresses (left) and ``busy'' addresses handling more than 100  transactions (right). Note the log scale.} 
\label{fig:reuse}
\end{figure}

\subsection{Insights into the Token Ecosystem}

The automatic detection of token systems allows us to shed more light into the token ecosystem. Looking at bytecode reuse, for instance, puts the headline numbers into perspective and informs us about the actual amount of innovation happening in the ICO community. To this end, Figure~\ref{fig:reuse} connects our technical level of analysis (bytecode instances) to the publicly visible level of addresses hosting token systems. The most frequently deployed bytecode instance of a token system is a standard template by \emph{ConsenSys}.\footnote{\url{https://github.com/ConsenSys/Token-Factory/blob/master/contracts/HumanStandardToken.sol}} It has been deployed \num{8729} times to the \ET blockchain. \num{298} of these deployments have processed more than 100 transactions. Altogether 49 bytecode instances have been deployed more than 100 times, and \num{16} bytecode instances have 10 or more ``busy'' deployments.\footnote{Note that ``busy'' is similar to our notion of \emph{active}, however on the level of addresses rather than bytecode instances.} These figures give some early intuition, but likely underestimate the extent of code reuse as trivial modifications of template code (or the output of token factories that deploy polymorphic code) are not consolidated.


\section{Related Work}
\label{sec:related_work}


As we are not the first to systematically analyze \SCS on \ET or to study tokens on the \EP, we summarize prior art by topic area.

\textbf{Mapping the Smart Contract Ecosystem:}
Using source code provided by Etherscan, Bartoletti and Pompianu~\cite{bartoletti2017empirical} manually classify \num{811} \SCS by application domain (\eg financial, gaming, notary) and identify typical design patterns.
Norvill \EA\cite{norvill2017automated} propose unsupervised clustering to group \num{936} \SCS on the \ET blockchain.
Zhou \EA\cite{zhou2018erays} develop Erays a Ethereum reverse engineering tool that lifts EVM bytecode to a human readable pseudocode representation, for futher inspection. They conduct four case studies to show the effectiveness of the approach. 

\textbf{Vulnerability Detection in Smart Contracts:}
Luu \EA\cite{luu2016making} execute \num{19366} \SCS symbolically with the intention to uncover security vulnerabilities, which they find in \num{8833} cases.
Tsankov \EA\cite{gervais} build \textsc{Securify}, a symbolic execution framework to uncover security problems in \SCS. Security patterns are specified in a domain-specific language based on Datalog.
Nikolic \EA\cite{nikolic2018finding} study so-called trace vulnerabilities that manifest after multiple runs of a program. Introducing \textsc{Maian}, a symbolic execution framework to reason about trace properties, they identify \num{3686} vulnerable smart contracts.
Brent \EA\cite{brent2018vandal} present \textsc{Vandal}, a smart contract security analysis framework. It uses 
a Datalog-based language tailored to describe static analysis checks.
%

\textbf{Token Systems:}
Somin \EA\cite{somin2018social} study network properties of token trades and show that the degree distribution has power-law properties. They use a simple token detection method based on \ERC events generated at runtime, therefore relying on the standard compliance of the contracts.
%
Etherscan identifies token systems using a signature-based approach~\cite{EtherscanHelp}. However, the details of the method are proprietary and thus not available for replicable science. Etherscan's headline numbers count addresses with code, not bytecode instances.

\textbf{Symbolic Execution:}
%
Symbolic execution is a very mature discipline as witnessed by the number of literature surveys published. For instance,
Baloni \EA\cite{baldoni2018survey} provide an overview of the main ideas and challenges in symbolic execution.
P{\u{a}}s{\u{a}}reanu \EA\cite{puasuareanu2009survey} offer a survey of trends in symbolic execution research and applications with special focus on test generation and program analysis.
%
Person \EA\cite{person2008differential} introduce differential symbolic execution to calculate behavioral differences between versions of programs or methods.

\textbf{Malware Detection:}
The prime application of symbolic execution in systems security is malware analysis.
Luo \EA\cite{7823022} use symbolic execution compare code based on behavior. 
Christodorescu \EA\cite{1425057} have developed a semantics-aware malware detection framework that uses templates to specify malicious patterns. 

\textbf{Financial Regulation:} 
We are not aware of symbolic execution in tools that support financial authorities in their monitoring and supervision tasks, although some applications stand to reason given the prevalence of algorithmic trading.

%

In contrast to the above-mentioned work on smart contracts and symbolic execution, we do not aim at generating test cases or show the absence of certain conditions in programs, \eg integer overflows.
We apply symbolic execution to explore all paths through a program and analyze whether that program can be classified based on a given structure, or the presence of certain behavior.

\section{Conclusion and Future Work}
\label{sec:conclusion}

The idea of this work is to detect \ET token systems based on behavioral patterns. We have presented a method and evaluated it as effective using curated ground truth data and a reference method based on signatures.

Both methods have specific advantages. The signature-based approach is simple, but limited to standard-compliant token systems. It is easy to defeat detection by slightly deviating from the standard. The method bears a false positive risk in case of factory contracts or dead code. Quantifying this risk is left as future work.
The method can be improved by taking data flow into account.

The behavior-based method does not depend on standard-compliance. It is robust against reordering of parameters or renaming of functions. To which extent it can deal with sophisticated obfuscation is left for future work. The effectiveness of this method demonstrates that symbolic execution is practical on \ET.

Both methods fail if the detectable pattern spans over more than address. If this limitation becomes problematic in practice, the use of concolic execution \cite{baldoni2018survey} in conjunction with the current blockchain state is a way to overcome the locality.

In particular the behavior-based method is hand-crafted to the application of token detection. A direction of future work is to generalize the approach by building a domain-specific language in which behavioral patterns can be specified on a high level of abstraction. 
This would facilitate extensions of our approach to detect other kinds of behavior, such as smart contracts implementing non-fungible tokens, decentralized exchanges, or gambling services. Evaluating the transactions between the so-identified services would provide the necessary information to draw a map of the Ethereum ecosystem.



\section*{Acknowledgments}
We like to thank ConsenSys for the work on \myth.
This work has received funding from the European Union's Horizon 2020 research and innovation programme under grant agreement No.\ 740558.



\bibliographystyle{splncs03}
\bibliography{paper}

\appendix

\clearpage
\section{Supplemental Result Tables}
\label{app:tabs}

\begin{table}
  \centering
  \caption{Recall of signature- and behavior-based detection methods against our GTD with lower signature threshold ($\ge 3$).}
  \label{tab:detection_gt_tresh}
  \DetectTable{\num{88.75}\% (\num{513})}{\num{0.00}\% (\num{0})}{\num{88.75}\% (\num{513})}
              {\num{11.25}\% (\num{65})}{\num{0.00}\% (\num{0})}{\num{11.25}\% (\num{65})}
              {\num{100.00}\% (\num{578})}{\num{0.00}\% (\num{0})}{\num{100.00}\% (\num{578})}
\end{table}

\begin{table}
  \centering
  \caption{Comparison of signature- and behavior-based detection methods on all bytecode instances with lower signature threshold ($\ge 3$).}
  \label{tab:detection_thresh}
  \DetectTable{\num{36.06}\% (\num{40346})}{\num{1.58}\% (\num{1772})}{\num{37.65}\% (\num{42118})}
              {\num{9.36}\% (\num{10473})}{\num{52.99}\% (\num{59291})}{\num{62.35}\% (\num{69764})}
              {\num{45.42}\% (\num{50819})}{\num{54.58}\% (\num{61063})}{\num{100.00}\% (\num{111882})}
\end{table}

\begin{table}[ht!]
  \centering
  \caption{Call instructions statistics for different bytecode subsets (mean / median).}
  \label{tab:avg_calls}
  \begin{tabular}{ C | C | C | C } 
    \textbf{Subset} &  \instruction{callcode} & \instruction{call} & \instruction{delegatecall} \\
    \hline
    - &  (\num{0.05} / \num{0}) & (\num{4.08} / \num{1}) & (\num{0.56} / \num{0}) \\   
    \SymOrSig &  (\num{0.05} / \num{0}) & (\num{2.91} / \num{1}) & (\num{0.10} / \num{0}) \\ %
    \SymAndSig &  (\num{0.05} / \num{0}) & (\num{1.34} / \num{1}) & (\num{0.70} / \num{0}) \\ %
    \rowcolor{lightgray!60}
    \nSymAndSig &  (\num{0.05} / \num{0}) & \textbf{(\num{9.06} / \num{6})} & (\num{0.20} / \num{0}) \\ 
    \SymAndnSig &  (\num{0.01} / \num{0}) & (\num{2.20} / \num{0}) & (\num{0.04} / \num{0}) \\ 
  \end{tabular}
\end{table}

\section{ERC-20 Interface Specification}
\label{app:erc}
\begin{minipage}{\linewidth} 
\begin{lstlisting}[language=Solidity, caption={\ERC interface in Solidity.},label={lst:ERC20}]
contract ERC20Interface {
    // Function Signatures
    function totalSupply() public constant returns (uint);
    function balanceOf(address tokenOwner)
        public constant returns (uint balance);
    function allowance(address tokenOwner, address spender) 
        public constant returns (uint remaining);
    function transfer(address to, uint tokens)
        public returns (bool success);
    function approve(address spender, uint tokens)
        public returns (bool success);
    function transferFrom(address from, address to, uint tokens)
        public returns (bool success);
    // Events
    event Transfer(address indexed from,
                   address indexed to,
                   uint tokens);
    event Approval(address indexed tokenOwner,
                   address indexed spender,
                   uint tokens);
}
\end{lstlisting}
\end{minipage}
\section{Documentation of Manual Inspections}
\label{app:manual}
\begin{table}[h!]
  \centering
  \caption{Five smart contracts of the GTD missed by the signature-based but found by the behavior-based method.}
  \label{tab:gt_sym_but_not_sig}
  \begin{tabular}{ m{2.3cm} | m{4.5cm} | m{3.cm} | c }
    \textbf{Name} & \textbf{Address} &  \textbf{Code Hash} & \multicolumn{1}{p{1.8cm}}{\centering \textbf{\# \ERC} \\ \textbf{Functions}}\\
    \hline
    \textit{LatiumX} & \hexitl{4.5cm}{2f85e502a988af76f7ee6d83b7db8d6c0a823bf9} & \hexitl{3cm}{f30b6028435e4e4dd4e2efaa1a30c6421a747c960f8362bbf627495b1529c685} & \num{3} \\
    \textit{Pylon} & \hexitl{4.5cm}{7703c35cffdc5cda8d27aa3df2f9ba6964544b6e} &\hexitl{3cm}{96858625adfad966b1a7037fdff791609039516f876ee7928262d84f75b86519} & \num{3} \\
    \textit{Minereum} & \hexitl{4.5cm}{1a95b271b0535d15fa49932daba31ba612b52946} & \hexitl{3cm}{65d59c447f7cbb17364dc29c7c6f1eef2805459c751c32d95f0cd33f963f220c} & \num{3}\\
    \textit{All Sports Coin} & \hexitl{4.5cm}{2d0e95bd4795d7ace0da3c0ff7b706a5970eb9d3} & \hexitl{3cm}{1c57e11bbd6e7628f7ec535743c2d13631f2dc7f90b166a1f949c5e4743a53d6} & \num{3} \\
    \textit{Golem} & \hexitl{4.5cm}{a74476443119a942de498590fe1f2454d7d4ac0d} & \hexitl{3cm}{35e72568bdaa9762a780d3b476a8a4dbf5fbe9b953720ce28d1dcdd9e889f95b} & \num{3}\\
  \end{tabular}
\end{table}

\begin{table}[h!]
  \centering
  \caption{Random sample of 20 smart contracts in the GTD missed by the behavior-based but found by signature-based method.}
  \label{tab:gt_sig_but_not_sym}
  \begin{tabular}{ m{3.5cm} | m{4.5cm} | m{3cm} }
    

    \multicolumn{2}{l}{\textsc{Delegation of Bookkeeping}} \\
    \cline{1-3}
    & \textbf{Address} &  \textbf{Code Hash}\\
     \textit{EmphyCoin} & \hexitl{4.5cm}{50ee674689d75c0f88e8f83cfe8c4b69e8fd590d} & \hexitl{3cm}{19780d1f0151fc0507cd9643a10c44e6b53ef8f40966553cd58d4dac71022cb7}  \\ 
     \textit{Digix Gold Token} & \hexitl{4.5cm}{4f3afec4e5a3f2a6a1a411def7d7dfe50ee057bf} & \hexitl{3cm}{941fab0f7c20663cb12cc9664b6779b0f093f320077b6215aa89ea4f24a49ed1}  \\ 

     \textit{FunFair} & \hexitl{4.5cm}{419d0d8bdd9af5e606ae2232ed285aff190e711b} & \hexitl{3cm}{e29653f94e73367250592ad244f8f9d52aef48b54be93251bed8f85a0d7953fb} \\ 

     \textit{Education} & \hexitl{4.5cm}{5b26c5d0772e5bbac8b3182ae9a13f9bb2d03765} & \hexitl{3cm}{e359bf40848dc14bb31d3aef623780b186ef2ab2ef863e6802a3c14c87a5c853} \\ 

     \textit{Devery.io} & \hexitl{4.5cm}{923108a439c4e8c2315c4f6521e5ce95b44e9b4c} & \hexitl{3cm}{6b8bff0af605183f68ad25faeb0509bf4dd17ddfeca37b96ddb541c284ce6c17} \\ 

      \textit{UniBright} & \hexitl{4.5cm}{8400d94a5cb0fa0d041a3788e395285d61c9ee5e} & \hexitl{3cm}{3058c20470fb76bfcefe77820cf8a639995227042c8ed3dea6b41d4627eca4ba} \\ 

     \multicolumn{2}{l}{\textsc{}} \\[5pt]

    \multicolumn{2}{l}{\textsc{Violation of Definition~\ref{def:TOK}}} \\
    \cline{1-3}
    & \textbf{Address} &  \textbf{Code Hash}\\

     \textit{Ethbits} & \hexitl{4.5cm}{1b9743f556d65e757c4c650b4555baf354cb8bd3} &\hexitl{3cm}{d3f5162252947ea6e46434553344ad4b4ffc147f99ceffe7e3690fdc5b046e61}   \\ 
     \textit{Aston X} & \hexitl{4.5cm}{1a0f2ab46ec630f9fd638029027b552afa64b94c} & \hexitl{3cm}{c2b817789336cd01f6bf9ad1650e585010bfd94d58a074ff9db9e469a5a76f74}  \\ 
     \textit{Sharpe Platform Token} & \hexitl{4.5cm}{ef2463099360a085f1f10b076ed72ef625497a06} & \hexitl{3cm}{e0e29e2655db0699fd15e0953932a4bd208253ac780f5df946dba7c545e03525} \\ 

     \textit{FundRequest} & \hexitl{4.5cm}{4df47b4969b2911c966506e3592c41389493953b} & \hexitl{3cm}{519dc5c0384bf7599845827c4d383d08e92819d5ec9d5372f145fec0ee2b0ead} \\ 

     \textit{SwarmCity} & \hexitl{4.5cm}{b9e7f8568e08d5659f5d29c4997173d84cdf2607} & \hexitl{3cm}{88b20869ae32e8f63a0cd81e6188e7b2accf52b7c31b9d2cbdc4e0c4ae1e72b6} \\ 

     \textit{Mothership} & \hexitl{4.5cm}{68aa3f232da9bdc2343465545794ef3eea5209bd} & \hexitl{3cm}{63e44909ce9322049ef0d36d98579cd56d813299d26734e429b0d16b791be7dd} \\ 

     \textit{Ethfinex Nectar Token} & \hexitl{4.5cm}{cc80c051057b774cd75067dc48f8987c4eb97a5e} & \hexitl{3cm}{5c7c39e24430f2b39ef45cd790e7286b54654cc4d13078ef412a23cd0c62687a} \\

     \textit{DaTa eXchange Token} & \hexitl{4.5cm}{765f0c16d1ddc279295c1a7c24b0883f62d33f75} & \hexitl{3cm}{c4bfdc9026f14edba3e3b1469c437fbc58c0a93235276e0abe60dd51d55368aa} \\ 

      \textit{Swarm Fund} & \hexitl{4.5cm}{9e88613418cf03dca54d6a2cf6ad934a78c7a17a} & \hexitl{3cm}{56dd7cb818b4347196ffb5be65bd66d5315646694f955d0e5cbf82fc3c9fcc28} \\

      \textit{MakerDAO} & \hexitl{4.5cm}{9f8f72aa9304c8b593d555f12ef6589cc3a579a2} & \hexitl{3cm}{e69355035f779ef6be33f5d3730a6da6eeae0325db58802a1b144c2cf2ffb3ee} \\ 

     \multicolumn{2}{l}{\textsc{}} \\[5pt]

   \multicolumn{2}{l}{\textsc{Limitations of Symbolic Execution and Taint Analysis}} \\
    \cline{1-3}
    & \textbf{Address} &  \textbf{Code Hash}\\
     \textit{Storiqa} & \hexitl{4.5cm}{5c3a228510d246b78a3765c20221cbf3082b44a4} & \hexitl{3cm}{93be59026507e637f56545daaf9f1a4cf59261d3cdee65bcf993204eeba51815} \\ 

     \textit{LocalCoinSwap Cr.} & \hexitl{4.5cm}{aa19961b6b858d9f18a115f25aa1d98abc1fdba8} & \hexitl{3cm}{88b9c793a727ab0fade0caab1372f789e1a2067e2831288f04a66b48d1181352} \\ 

      \textit{LOCIcoin} & \hexitl{4.5cm}{9c23d67aea7b95d80942e3836bcdf7e708a747c2} & \hexitl{3cm}{9488b89a5ee6e9d0bf26a6a010d820e3a157726e2dfc65654cf4cf6c1ef08366} \\ 

     \textit{TrueUSD} & \hexitl{4.5cm}{8dd5fbce2f6a956c3022ba3663759011dd51e73e} & \hexitl{3cm}{f447f893b44fdd0a44bc117f600a401de8a153ddb2bb45de3b697e17503c211c} \\ 
  \end{tabular}
\end{table}

\end{document}